\RequirePackage{fix-cm}
\documentclass[twocolumn]{svjour3}          
\smartqed  
\usepackage{graphicx}
\usepackage{pgf}
\usepackage{amsmath}
\usepackage{amstext}
\usepackage{url}
\usepackage{cite}
\usepackage{enumitem}
\usepackage{siunitx}

\begin{document}

\title{Fermions meet two bosons - the heteronuclear Efimov effect revisited
}


\author{Binh Tran         \and
        Michael Rautenberg \and
        Manuel Gerken \and
        Eleonora Lippi \and
        Bing Zhu \and
        Juris Ulmanis \and
        Moritz Drescher \and
        Manfred Salmhofer \and
        Tilman Enss \and
        Matthias Weidem\"uller 
}


\institute{B. Tran \and Michael Rautenberg \and Manuel Gerken \and Eleonora Lippi \and Bing Zhu \and Juris Ulmanis \and Matthias Weidem\"uller\at
              Physikalisches Institut, Universit\"at Heidelberg, Im Neuenheimer Feld 226, 69120 Heidelberg, Germany  \\ \email{tran@physi.uni-heidelberg.de}
		\and Moritz Drescher \and Manfred Salmhofer \and Tilman Enss \at Institut f\"ur Theoretische Physik, Universit\"at Heidelberg, Philosophenweg 19, 69120 Heidelberg, Germany
		\and Bing Zhu \at Hefei National Laboratory for Physical Sciences at the Microscale and Department of Modern Physics,
		and CAS Center for Excellence and Synergetic Innovation Center in Quantum Information and Quantum Physics,
		University of Science and Technology of China, Hefei 230026, China
}

\date{Received: date / Accepted: date}

\maketitle

\begin{abstract}
In this article, we revisit the heteronuclear Efimov effect in a Bose-Fermi mixture with large mass difference in the Born-Oppenheimer picture. As a specific example, we consider the combination of bosonic $^{133}\mathrm{Cs}$ and fermionic $^6\mathrm{Li}$. In a system consisting of two heavy bosons and one light fermion, the fermion-mediated potential between the two heavy bosons gives rise to an infinite series of three-body bound states. The intraspecies scattering length determines the three-body parameter and the scaling factor between consecutive Efimov states. In a second scenario, we replace the single fermion by an entire Fermi Sea at zero temperature. The emerging interaction potential for the two bosons exhibits long-range oscillations leading to a weakening of the binding and a breakup of the infinite series of Efimov states. In this scenario, the binding energies follow a modified Efimov scaling law incorporating the Fermi momentum. The scaling factor between deeply bound states is governed by the intraspecies interaction, analogous to the Efimov states in vacuum. 

\end{abstract}

\section{Introduction}
\label{intro}

The late Mahir S. Hussein, to whom this special issue is dedicated, is best known for his ground-breaking contributions in nuclear physics, e.g., to the understanding of the fusion and break-up of loosely bound nuclei \cite{Canto2006}. From a very early stage on, Mahir also recognized the importance of ultracold atomic gases, in particular Bose-Einstein condensates, for reaching a deeper understanding of strongly-correlated many-body quantum systems. In particular, based on his solid background in nuclear physics, he was among the first to realize the immense potential of magnetically tuned Feshbach resonances for tuning effective interactions over a wide range \cite{Timmermans1999}. Therefore, it appeared appropriate to us to discuss one specific, rather spectacular example of the application of tunable interparticle interactions on the quantum level, namely the so-called Efimov effect. This effect, originally proposed in the context of nuclear binding \cite{Efimov1970}, nicely demonstrates the fruitful cross-fertilization of concepts between nuclear physics and ultracold atomic quantum gases, which was so dear to Mahir S. Hussein.

The unexpected emergence of an infinite series of bound three-particles states via resonant pairwise interaction was first predicted theoretically by Vitaly Efimov in 1970 \cite{Efimov1970}. Since then, the Efimov effect has become a prime example for studying universality in quantum few-body systems. In the universal regime, i.e. when the scattering length $a$ between two particles exceeds the characteristic range of interparticle interaction $r_0$, there exists a series of three-body bound states which follow a simple discrete scaling law. Remarkably, these three-body bound states can form, even though the two-body interactions are too weak to support a two-body bound state. Initially, Efimov's prediction raised serious doubts about the applicability of his concepts, but theorists trying to prove him wrong had to finally concede that he might actually be right. While Efimov proposed the effect to be observed in nuclear systems, such as in $^3\mathrm{He}^+$ or in the Hoyle state of $^{12}\mathrm{C}$, the existence of a series of three-bound states requires the two-body interactions to be close to resonance. It was only in 2006, 36 years after Efimov's theoretical prediction, that first experimental evidence was eventually found in a gas of ultracold $^{133}\mathrm{Cs}$ atoms \cite{Kraemer2006}. The precise tuning of two-body interactions from infinite repulsion to attraction via Feshbach resonances opened up unique opportunities to study the range of universality in the Efimov scenario. Following the investigations in homonuclear systems, the Efimov effect was later also observed in heteronuclear systems \cite{Barontini2009, Barontini2010, Bloom2013, Pires2014, Tung2014, Hu2014, Maier2015 } where, e.g., a large mass ratio \cite{Pires2014, Tung2014} leads to a denser Efimov spectrum which allowed for the observation of up to three consecutive Efimov resonances.

Besides its relevance in the field of few-body physics, the Efimov effect plays an important role in the understanding of many-body systems in the presence of three-body bound states. Similarly, one can consider the Efimov effect from a few-body perspective and study the influence of a surrounding many-body background to the trimer. The latter has been studied in different scenarios with either a Fermi Sea \cite{Nygaard2014, MacNeill2011, Sun2019, Sanayei2020} or a BEC \cite{Zinner2013, Naidon2018} serving as a background. The former has been studied in \cite{Levinsen2015, Sun2017, Sun2017b}, following recent experimental advances in creating Bose \cite{Scelle2013,Jorgensen2016, Hu2016, Yan2020, Skou2020} and Fermi polarons \cite{Schirotzek2009, Nascimbene2009, Kohstall2012, Koschorreck2012, Cetina2016, Scazza2017, Yan2019}. More details on different aspects of the Efimov effect are covered extensively in several reviews, which have appeared in the years after its first realization \cite{Braaten2006, Braaten2007, Ferlaino2011, Blume2012, Wang2013, Wang2015, Naidon2017}.

In this work, we will focus on the specific case of two heavy bosons and a light fermion, which allows to apply the Born-Oppenheimer (BO) approximation. We investigate two limiting cases of the Efimov scenario, first, in vacuum, and second, in the presence of a Fermi Sea. While the first case reproduces the well-known features of the Efimov effect, the second case provides novel insights serving as a precursor to understand effective interactions of Fermi polarons \cite{Enss2020}, i.e., strongly correlated impurities in a Fermi sea. As a specific example, we consider the heteronuclear mixture of bosonic $^{133}\mathrm{Cs}$ and fermionic $^6\mathrm{Li}$, for which experiments have been performed by us and others. Providing a simple and intuitive access to understand the Efimov effect in a mass-imbalanced system, the BO approximation does not only capture the existence of an infinite series of three-body bound states\cite{Fonseca1979, Braaten2006, Bhaduri2011,Petrov2012}, but can also account for intraspecies interactions via short-range van der Waals (vdW) potentials \cite{Wang2012}. Our paper is structured as follows: after giving a basic introduction to the Efimov scenario in Sec. \ref{sec:the_efimov_scenario}, we describe how we solve the Schrödinger equation by employing the BO approximation in Sec. \ref{sec: born-oppenheimer approximation}. In Sec. \ref{sec:two_cs_atoms_+_one_li_atom}, we present our results on the Efimov energy spectrum for a system consisting of two Cs atoms and one Li atom, taking into account finite Cs-Cs s-wave interaction.  We extend our results to a system of two Cs atoms immersed in a Fermi Sea (Sec. \ref{sec: two_cs_atoms_in_a_li_fermi_sea}), before concluding in Sec. \ref{sec:conclusions}.

\section{The Efimov Scenario}\label{sec:the_efimov_scenario}
\begin{figure}
	\centering
	\includegraphics[width=1.0\linewidth]{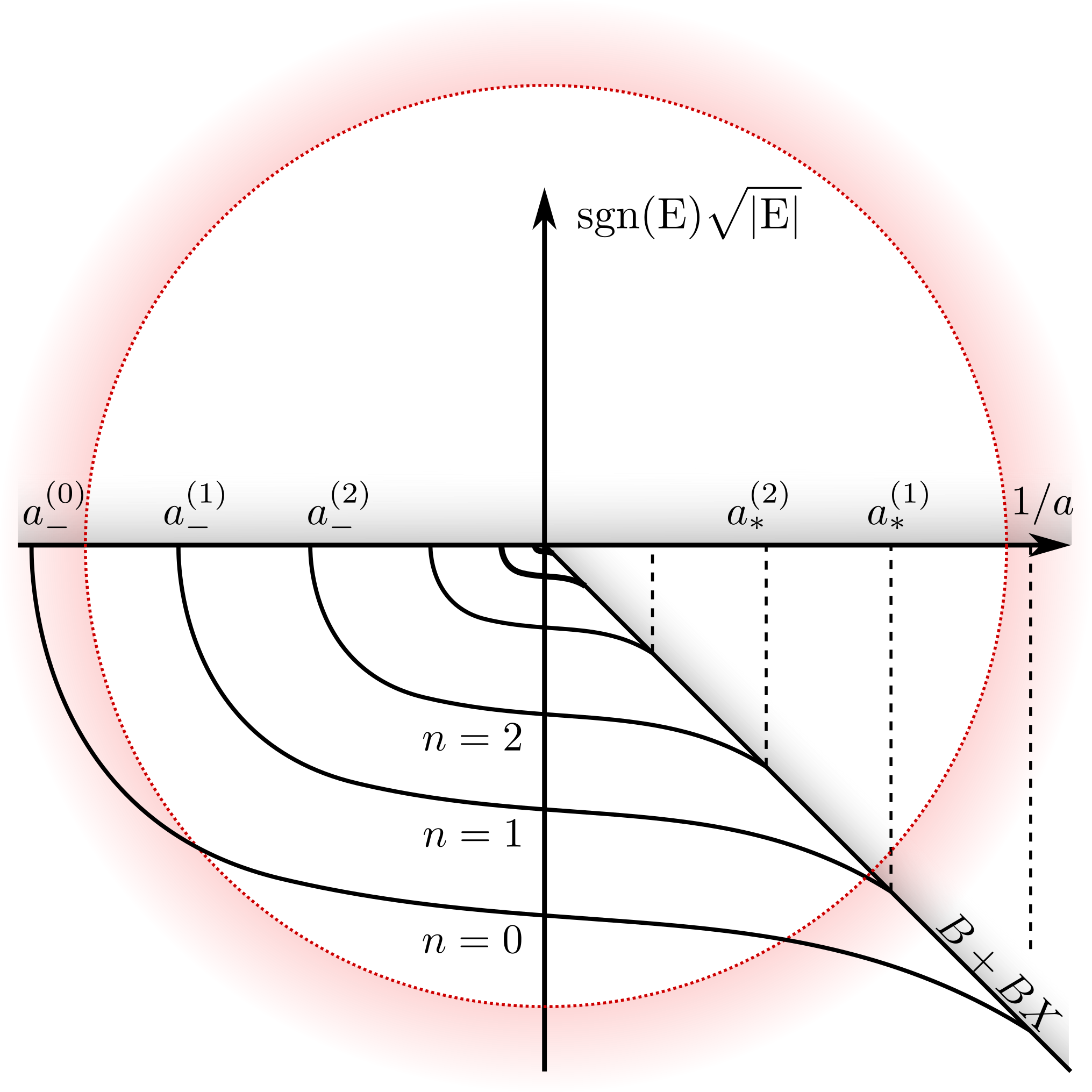}
	\caption{Heteronuclear Efimov Scenario for two identical bosons B and one distinguishable particle X. Shown are the few deepest bound Efimov trimers (solid lines) connecting the three-body dissociation threshold with the atom-dimer threshold $B + BX$. The values of the scattering lengths where the Efimov trimers cross the dissociation threshold are marked by $a_-^{(n)}$ and $a_*^{(n)}$ for $a<0$ and $a>0$, respectively. The intraspecies scattering length $a_{\mathrm{BB}}$ between the particles $B$ is resonant. Short-range two-body interactions (red shaded area) limit the discrete scaling law between Efimov trimers.}
	\label{fig:Efimov_Scenario}
\end{figure}
The Efimov scenario for the heteronuclear case can be visualized in an energy diagram (Fig. \ref{fig:Efimov_Scenario}) which is plotted against the inverse interspecies scattering length $1/a$ between the identical bosons B and the distinguishable particle X. For simplicity, we first consider the scenario in which the two bosons resonantly interact ($a_{\mathrm{BB}} \rightarrow \infty$). Then the energy diagram can be divided into three different regions representing the three-body scattering states, Efimov trimers, and atom-dimer states.  For positive energies $(E>0)$, the three atoms are unbound and possess a finite kinetic energy. For negative energies $(E<0)$, one needs to distinguish between the sign of the scattering length. On the positive scattering length side $(a>0)$, the system supports a weakly bound dimer state BX with an energy of $1/2\mu a^2$, where $\mu$ is the reduced mass. Above the atom-dimer threshold, the dimer state BX coexists with a free atom B. The Efimov trimers exist in the region below the three-body dissociation threshold at $a<0$ and the atom-dimer threshold at $a>0$. In this region, an infinite number of Efimov trimers with energies $E_n$ exist which cross the three-body dissociation threshold at values of $a_{-}^{(n)}$ and $a_{*}^{(n)}$ on the negative and positive scattering length side, respectively.
These crossings can be described by the discrete scaling laws
\begin{equation}
\begin{aligned}
	a_{-}^{(n+1)} &= \lambda a_{-}^{(n)},\\
	a_{*}^{(n+1)} &= \lambda a_{*}^{(n)},\\
	E_{n+1} &= \lambda^{-2} E_{n}
\end{aligned}
\label{eq:scaling}
\end{equation}
where the scaling factor $\lambda = e^{\pi/s_0}$ with the dimensionless parameter $s_0$ is dependent on the mass ratio, the number of resonant interactions and the quantum statistics of the particles \cite{Naidon2017}.

 In real systems, interactions have a finite range, given by non-zero interparticle distances in the underlying interaction potentials. Therefore, in the infinite progression of Efimov trimers a ground state has to be considered. This can be done by introducing a three-body parameter (3BP) which defines the position of the energy $E_0$ or the scattering length $a_{-}^{(0)}$ of the lowest Efimov trimer. The 3BP is determined by short-range two-body interactions and can be expressed in terms of vdW units (see Eq. \ref{eq:vdW}). As indicated in Fig. \ref{fig:Efimov_Scenario}, the universal Efimov scaling therefore only holds if $|a| \gg max(r_{vdW}^{BX}, r_{vdW}^{BB})$ due to finite range effects.

\section{Born-Oppenheimer approximation}\label{sec: born-oppenheimer approximation}

In the Born-Oppenheimer approximation, we can solve the three-body Schrödinger equation for two heavy bosons with mass $M$ and a light atom with mass $m$ in a two-step approach. We assume that the mass ratio is $M/m \gg 1$, such that the light atom immediately follows the motion of the heavy ones. In the first step, the Schrödinger equation is solved for the light atom with a potential created by the two heavy scatterers at fixed distances. The resulting energy serves as an interaction potential $V_{\mathrm{E}} (R)$, induced by the presence of the light particle, in the Schrödinger equation for the heavy bosons
\begin{align}
\left[-\frac{1}{M} \nabla^2_\textbf{R} + V_{\mathrm{BB}}(R) + V_{\mathrm{E}}(R)\right] \phi(\textbf{R}) = E \phi(\textbf{R})
\label{eq:schroedinger}
\end{align}
where $\textbf{R}$ denotes the distance between the two heavy bosons. 
The boson-boson interaction is modeled by a van der Waals potential with a hard core of the form \cite{Gribakin1993, Flambaum1999}
\begin{align}
V_{\mathrm{BB}}(R) = \begin{cases}\infty, & R <R_0\\-C_6/R^6, & R>R_0.\end{cases}
\label{eq:v_bb}
\end{align}
The $C_6$ coefficient allows us to naturally introduce the van der Waals radius $r_{\mathrm{vdW}}$ and energy $E_{\mathrm{vdW}}$ via \cite{Chin2010}

\begin{align}
	r_{\mathrm{vdW}} = \frac{1}{2} (M C_6)^{1/4}
	\label{eq:vdW}
\end{align}
and 
\begin{align}
	E_{\mathrm{vdW}} = \frac{1}{M r_{\mathrm{vdW}}^2}.
\end{align}
The cut-off radius $R_0$ in Eq. \ref{eq:v_bb} determines the 3BP and can be directly related to the boson-boson scattering length $a_{\mathrm{BB}}$ via \cite{Gribakin1993}
\begin{align}
\frac{N_{1/4} (2 r_{\mathrm{vdW}}/R^2_0)}{J_{1/4} (2 r_{\mathrm{vdW}}/R^2_0)} =  1-\sqrt{2}\frac{a_{\mathrm{BB}}}{r_{\mathrm{vdW}}}\frac{\Gamma(5/4)}{\Gamma(3/4)}
\label{eq:cut-off}
\end{align}
where $J_\nu(x)$ and $N_\nu(x)$ are Bessel functions of the first and second kind, respectively. As Eq. \ref{eq:cut-off} has more than one solution, the value of $R_0$ does not only determine $a_{BB}$, but also the number of bound states supported by the vdW potential $V_{\mathrm{BB}}(R)$.  However, we note that the exact number of dimer states is irrelevant for our purposes as it has no significant effect on the long-range part of the Efimov wavefunctions \cite{Wang2012}.

The induced interaction potential $V_E(R)$ possesses a symmetric and an antisymmetric solution \cite{Petrov2012}. In the limit of $a \rightarrow \infty$, only the former is relevant for the discussion of bound states. It reads

\begin{align}
	V_E(R) = -\frac{c^2}{2mR^2}
\end{align}
where $c \approx 0.567$ is the solution of $c = e^{-c}$ and is connected to the scaling factor $s_0$ via 
\begin{align}
	c = \sqrt{\frac{2m}{M} \left(s_0^2 + \frac{1}{4}\right)}.
\end{align}
With the interaction potentials $V_E(R)$ and $V_{\mathrm{BB}}(R)$ we can solve Eq. \ref{eq:schroedinger} in order to understand the Efimov scenario for a Cs-Cs-Li system and its scaling behavior taking into account boson-boson interactions.  

\section{Two bosons meet one fermion}\label{sec:two_cs_atoms_+_one_li_atom}
\begin{figure}
	\resizebox{\linewidth}{!}{\input{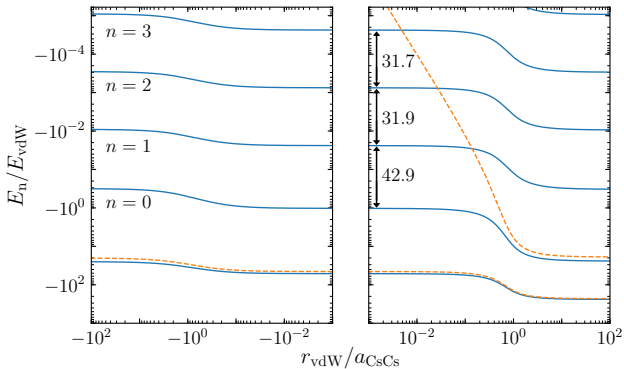}}
	\caption{Energy spectrum of a Cs-Cs-Li system at heteronuclear unitarity across a Cs-Cs resonance (blue solid lines) and the comparison to the energy spectrum of the two most weakly bound $\mathrm{Cs_2}$ dimers (orange dashed lines).  }
	\label{fig:binding_energy_efimov}
\end{figure}

In Fig. \ref{fig:binding_energy_efimov} we show the solution of Eq. \ref{eq:schroedinger} using the Li-Cs mass ratio of $M_{\mathrm{Cs}}/m_{\mathrm{Li}} = 22.1 $ and a short-range cutoff $R_0$ for which the energy spectrum supports up to two $\mathrm{Cs_2}$ dimer states. It is instructive to consider the two cases
\begin{enumerate}[label=\alph*)]
	\item only boson-boson interaction $V_{\mathrm{BB}}(R)$
	\item total potential $V_{\mathrm{E}}(R) + V_{\mathrm{BB}}(R).$
\end{enumerate}

For case a), i.e. pure two-body vdW interaction, we identify two weakly bound dimer states (orange dashed lines). The energy of the least bound state approaches the binding energy of the $\mathrm{Cs_2}$ dimer $E_b = 1/M a_{\mathrm{CsCs}}^2$ for positive and increasing $a_\mathrm{CsCs}$. The energy of the most deeply bound state crosses the Cs-Cs resonance and shows a step-like behavior around $a_\mathrm{CsCs} = r_{\mathrm{vdW}}$. This behavior marks the crossover between a vdW-dominated ($a_\mathrm{CsCs} < r_{\mathrm{vdW}}$) dimer and a halo state ($a_\mathrm{CsCs} > r_{\mathrm{vdW}}$). For case b) (blue lines), we find that the energy of the most deeply bound state closely follows case a), such that we can assign this state to the $\mathrm{Cs_2}$ dimer state. The next bound state $E_{n = 0}$ does not follow the respective $\mathrm{Cs_2}$ dimer state anymore, but persists also across the Cs-Cs resonance. This clearly shows the effect of the mediated interaction necessary to form Efimov trimer states. The following bound states with energies $E_{n \geq 1}$ correspond to the infinite progression of Efimov states. They show a gradual step around $a_\mathrm{CsCs} = r_{\mathrm{vdW}}$ which we assign to an overlap of the wavefunctions of the Efimov trimers with the $\mathrm{Cs_2}$ halo dimer. This crossover between states originating from a short-range molecular vdW potential and a long-range $\propto -1/R^2$ potential is one of the main results of the finite range BO approximation, providing an intuitive access to the three-body problem. From the calculated energy spectrum, we can additionally extract the scaling factor between adjacent energy levels. For $\lambda_{n}^2 := E_n/E_{n-1}$  the scaling factor between the two most deeply bound states amounts to $\lambda_{n=1}^2 = 42.9$ at resonance. When crossing the resonance towards negative $\mathrm{a_{CsCs}}$, the scaling factor approaches the universal value of \\ $\lambda_{n \rightarrow \infty}^2 = (5.63)^2 = 31.7$ for a pure $\propto -1/R^2$ potential. The deviation close to resonance can again be explained by the existence of the weakly bound $\mathrm{Cs_2}$ dimer state. We find that already the second deepest scaling factor $\lambda_{n=2}^2 \approx 31.9$ is close to the universal value.

\section{Two bosons meet the Fermi Sea}\label{sec: two_cs_atoms_in_a_li_fermi_sea}

Let us now consider two heavy bosons immersed in a Fermi Sea. In the BO approximation the effective Li-mediated potential between the two bosons can be calculated by \cite{Nishida2009}:
\begin{align}
V_{\mathrm{eff}}(R) = \Delta E(R) - \Delta E(R \rightarrow \infty)
\end{align}
Here $\Delta E(R)$ denotes the energy reduction of the interacting system compared to the free system
\begin{align}
\Delta E(R) = - \frac{\kappa^2_{+} + \kappa^2_{-}}{2m} - \int_{0}^{k_F}dk k \frac{\delta_{+} (k) + \delta_{-}(k)}{m \pi}
\end{align}
with the bound state wavevectors and scattering phase shifts described by $\kappa_{\pm}$ and $\delta_{\pm}$, respectively \cite{Nishida2009}. In the asymptotic limit, the energy reduction $\Delta E(R \rightarrow \infty)$ is equivalent to the chemical potential of two free, heavy atoms in a Fermi Sea \cite{Combescot2007}. The length scale of the Fermi Sea is governed by the Fermi wavevector $k_F$ which is related to the atomic density $n$ via $k_F = (6 \pi^2 n)^{1/3}$.
\begin{figure}
	\resizebox{\linewidth}{!}{\input{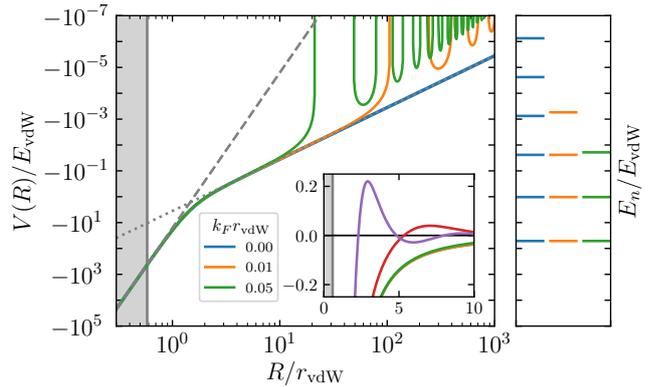}}
	\caption{Left panel: interaction potential between two Cs atoms consisting of a short-range vdW potential with $a_{\mathrm{CsCs}} = a_{\mathrm{LiCs}} =  \infty$ (grey dashed line) and a Li-mediated long-range potential. For comparison the $\propto -1/R^2$ potential is shown as a dotted line. In the inset the total potential is plotted also for the values $k_F r_{\mathrm{vdW}} = 0.2 $ (red line) and $k_F r_{\mathrm{vdW}} = 0.5$ (purple line). Right panel: corresponding binding energies for different $k_F$.}
	\label{fig:eff_potential}
\end{figure}
In Fig. \ref{fig:eff_potential} (left panel) we plot the total potential $V(R) = V_{\mathrm{eff}}(R) + V_{BB} (R)$ for which we solve the Schrödinger equation for unitarity $a_{\mathrm{LiCs}} = a_{\mathrm{CsCs}} = \infty$. The grey shaded area marks the hard wall for $R < R_0$. If we set $k_F r_\mathrm{vdW}= 0$ (blue line), we recover the potential from the Efimov scenario consisting of a short-range $\propto -1/R^6$ vdW potential (grey dashed line) and a long-range Efimov potential $\propto -1/R^2$ (grey dotted line). For increasing $k_F$, the effective potential starts to grow a repulsive barrier around $R \approx k_F^{-1}$ showing damped oscillations. In the inset of Fig. \ref{fig:eff_potential}, this behavior can be seen more clearly for even larger values of $k_F$. The form of the potential is reminiscent of the form of Friedel oscillations \cite{Friedel1952} which arise due to the sharp edge of the Fermi distribution. 
In fact, in the limit of $a_\mathrm{LiCs} \ll R, k_F^{-1}$, $V_\mathrm{eff}(R)$ takes the same form  of the Friedel oscillations or of the RKKY interaction \cite{Ruderman1954} in the context of magnetic interactions.

We now use the effective potential $V_{\mathrm{eff}} (R)$ to replace the Efimov potential $V_E (R)$ in Eq. \ref{eq:schroedinger} and solve for the eigenenergies of the system (right panel of Fig. \ref{fig:eff_potential}). We note that the chosen values of $k_F r_{\mathrm{vdW}} = 0.01$ and $k_F r_{\mathrm{vdW}} = 0.05$ correspond to densities between $10^{11}$ \si{cm^{-3}} and $10^{13}$ \si{cm^{-3}}. For $k_F r_{\mathrm{vdW}} = 0.01$, the $\mathrm{Cs_2}$ dimer state and the  deepest two Efimov states remain unchanged. For the following state, we see that the effective potential starts to deviate from the Efimov potential around $R \approx \SI{70}{r_{vdW}} $ leading to a smaller binding energy. As the potential gets more repulsive for increasing $R$, the formation of bound states is completely suppressed. Similarly, for $k_F r_{\mathrm{vdW}} = 0.05$, the weakening of the binding can be understood from a deviation of the Efimov potential around $R \approx \SI{10}{r_{vdW}} $ and the suppression of the infinite series of Efimov states begins one state earlier.

\begin{figure}
	\resizebox{\linewidth}{!}{\input{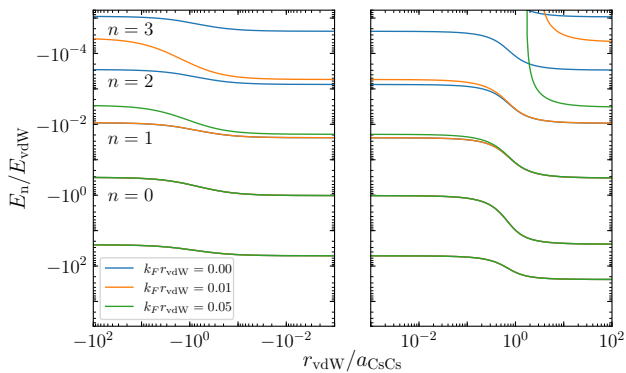}}
	\caption{Energy spectrum of two heavy Cs atoms in a Li Fermi Sea ($k_F r_{\mathrm{vdW}} >0 $) compared to the Cs-Cs-Li system ($k_F r_{\mathrm{vdW}} = 0 $, blue solid lines) in the unitarity limit $a_{\mathrm{LiCs}} \rightarrow \infty$. Finite densities of the Fermi Sea lead to a suppression of bound states and to the breakdown of the discrete scaling behavior of Eq.\ref{eq:scaling}.}
	\label{fig:binding_energy}
\end{figure}

Analogous to the case of two Cs atoms and one Li atom, we want to investigate the role of the intraspecies interaction, as shown in Fig. \ref{fig:binding_energy}. The energies in the case $k_F r_{\mathrm{vdW}} = 0$ again represent the previous Efimov bound state energies. Introducing a Fermi Sea to the system ($k_F r_{vdW} = 0.05$) we see that, on the positive scattering length side, the $n=1$ state shows first deviations from the Efimov scaling when $a_{\mathrm{CsCs}}$ takes values larger than $r_{\mathrm{vdW}}$ at the step going from the vdW-dominated to the long-range regime. Crossing the resonance towards negative scattering lengths, this deviation gets larger again at the step $-a_{\mathrm{CsCs}} \approx r_{\mathrm{vdW}}$. Following the energy line further to the positve scattering length side, the state rapidly dissociates before it reaches the long-range regime. In the same way, with a smaller density of the Fermi Sea of $k_F r_{vdW} = 0.01$, the system supports one more bound state before the binding energy vanishes.

\begin{figure}[t]
	\includegraphics[width=1.0\linewidth]{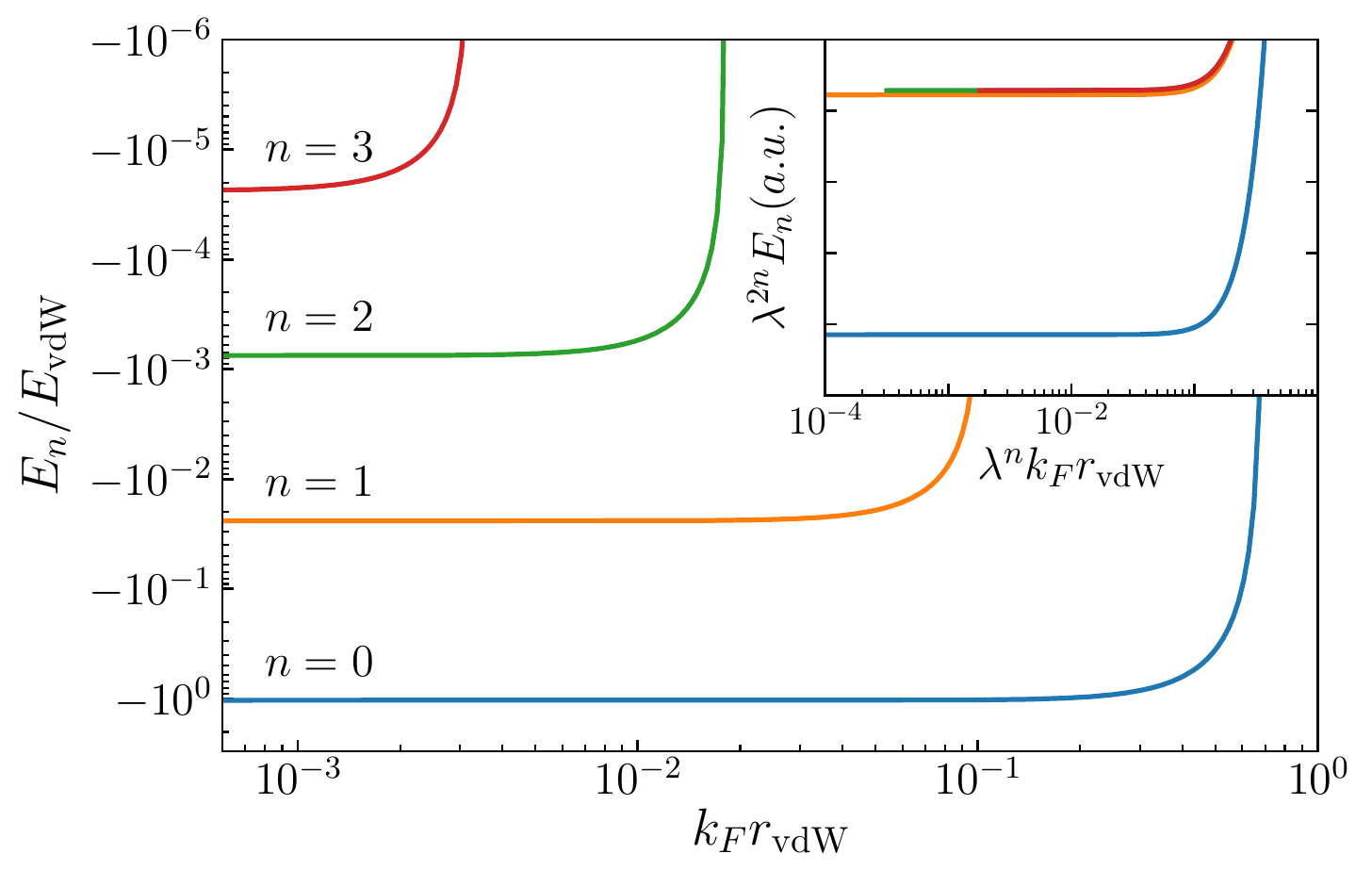}
	\caption{Dependence of the binding energies on the Fermi wavevector $k_F$ for intraspecies resonance. Inset: The discrete scaling of Eq. \ref{eq:scaling_law_fermi_sea} does not hold for the ground state (blue line), while high-lying bound states $n\geq 2$ fall on the same curve confirming the scaling.}
	\label{fig:binding_energy_cut}
\end{figure}

We calculate the density dependence of the binding energies (Fig. \ref{fig:binding_energy_cut}) and find that they remain nearly constant before they rapidly go to the continuum. The number of bound states is given by the Fermi wavevector $k_F$ and the shape of the lines suggest a similar scaling as we have seen in Fig. \ref{fig:Efimov_Scenario} following Eq. \ref{eq:scaling}. 
Indeed, in the presence of the Fermi Sea, the two heavy bosons follow a new discrete scaling law including the additional length scale $k_F$ \cite{Nygaard2014, Sun2019}:
\begin{equation}
\begin{aligned}
	a_{-}^{(n+1)}(k_F) &= \lambda a_{-}^{(n)}(\lambda k_F),\\
	E_{n+1} (k_F, a)&= \lambda^{-2} E_{n} (\lambda k_F, \lambda^{-1} a)
\end{aligned}
\label{eq:scaling_law_fermi_sea}
\end{equation}
In the case of finite intraspecies interactions we find that this scaling is fulfilled for high-lying bound states $n>1$ (inset of Fig. \ref{fig:binding_energy_cut}). Analogous to the scenario of two Cs atoms and one Li atom in Sec. \ref{sec:two_cs_atoms_+_one_li_atom}, the first excited bound state ($n=1$, orange line) shows only a small deviation from the scaling, whereas the scaling is broken for the ground state due to finite range effects. 

\section{Conclusions}\label{sec:conclusions}
In summary, we have calculated the binding energies of a Cs-Cs-Li system and of a system of two Cs atoms in a Li Fermi Sea and studied the influence of the intraspecies scattering length using the BO approximation. In the Cs-Cs-Li system, the intraspecies interaction leads to a step-like behavior in the energy spectrum and the existence of weakly bound $\mathrm{Cs}_2$ dimers influence the scaling factor. 
Immersing the two Cs atoms in a Li Fermi Sea suppresses the formation of bound states for sufficiently high $k_F$ and breaks the discrete Efimov scaling law. Instead, a new scaling law can be formulated which takes into account the wavevector $k_F$. This additional length scale of the Fermi Sea may also be used to define a new window of universality which is not only determined by short-range interactions, but also by the Fermi wavevector. In an experiment, the shifted position of the Efimov states in the Fermi Sea, which can also be interpreted as bipolaronic states, may be observed by means of three-body loss measurements, similar to the previous Efimov experiments \cite{Ulmanis2016} while now lower temperatures and higher densities are required.  The influence of the intraspecies scattering length can be studied in the Li-Cs system which features two interspecies Feshbach resonances around \SI{843}{G} and \SI{889}{G} with negative and positive sign of the intraspecies scattering length, respectively.   
However, we note that our simple BO approximation does only provide qualitative results. For the Cs-Cs-Li system, more quantitative results beyond the BO approximation can be obtained by means of a spinless vdW theory \cite{Haefner2017} where the three-body problem is solved in the hyperspherical formalism with two-body interactions modeled by a single channel Lennard Jones potential. 
Also finite temperature effects as well as scattering of trimers by the Fermi Sea which may lead to the excitation of particle-hole pairs and a change of the effective interaction potential \cite{MacNeill2011} have to be considered for a more realistic description of the system.

\begin{acknowledgements}
This work is supported by the Deutsche Forschungsgemeinschaft (DFG, German Research Foundation) - Project-ID 273811115 - SFB 1225 ISOQUANT and by DFG under Germany's Excellence Strategy EXC-2181/1 - 390900948 (Heidelberg STRUCTURES Excellence Cluster). E.L. acknowledges support by the IMPRS-QD.
\end{acknowledgements}

%
%

\bibliographystyle{spphys}       

\bibliography{references}


%
%

\end{document}